\documentclass[aps,prb,showpacs,twocolumn,superscriptaddress]{revtex4}
\usepackage{amsmath,graphicx}
\usepackage{epsfig}
\usepackage[next]{inputenc}

\def\be{\begin{equation}}
\def\ee{\end{equation}}

\def\bi{\begin{itemize}}
\def\ei{\end{itemize}}
\def\bn{\begin{enumerate}}
\def\en{\end{enumerate}}
\def\bea{\begin{eqnarray}}
\def\eea{\end{eqnarray}}
\def\no{\nonumber}
\def\ba{\begin{array}}
\def\ea{\end{array}}
\def\bd{\begin{displaymath}}
\def\ed{\end{displaymath}}
\def\la{\langle}
\def\ra{\rangle}

\begin{document}


\title{Phase diagram of the one dimensional anisotropic Kondo-necklace model}

\author{S. Mahmoudian}
\affiliation{Physics Department, Sharif University of Technology, Tehran 11155-9161, Iran}
\author{A. Langari}
\email[]{langari@sharif.edu}
\homepage[]{http://spin.cscm.ir}
\affiliation{Physics Department, Sharif University of Technology, Tehran 11155-9161, Iran}

\begin{abstract}
The one dimensional anisotropic Kondo-necklace model has been studied by several methods.
It is shown that a mean field approach fails to gain the correct phase diagram for the
Ising type anisotropy.
We then applied the spin wave theory which is justified for the anisotropic 
case. We have derived the phase diagram between the antiferromagnetic long range order 
and the Kondo singlet phases. We have found that the exchange interaction ($J$) between
the itinerant spins and local ones enhances the quantum fluctuations around the classical
long range antiferromagnetic order and finally destroy the ordered phase at the critical 
value, $J_c$.  
Moreover, our results show that the onset of anisotropy in the
XY term of the itinerant interactions develops the antiferromagnetic order for $J<J_c$. 
This is in agreement with the qualitative feature which we expect from the symmetry of 
the anisotropic XY interaction.
We have justified our results by the numerical Lanczos method where the structure
factor at the antiferromagnetic wave vector diverges as the size of system goes to infinity.
\end{abstract}
\date{\today}


\maketitle


\section{Introduction}
Quantum phase transitions between
Kondo singlet and antiferromagnetically ordered states such as found in heavy fermion 
compounds have been attracted much research interest recently \cite{Continentino05,Vojta03,Fulde06}. 
It is therefore of great importance to understand the approach to
the quantum critical region within suitable theoretical models. The
most important among them is the Kondo lattice model consisting of a
free conduction band and an on-site antiferromagnetic Kondo interaction which favors
non-magnetic singlet formation. In  the second order it also leads to
the effective inter-site Ruderman-Kittel-Kasuya-Yosida (RKKY) 
interactions which favor magnetic
order. Their competition leads to the appearance of the quantum critical point. In such a
picture only spin degrees of freedom are involved in
the quantum phase transition. Therefore the Kondo lattice model may be
replaced by a simpler model where the itinerant hopping part is
simulated by an inter-site interaction of the itinerant spins. 
Kondo necklace (KN) model has been originally proposed by Doniach \cite{Doniach77} for the one-dimensional case as a simplified version of the itinerant Kondo lattice (KL) model \cite{Tsunetsugu97}. Thereby the kinetic energy of conduction electrons is replaced by an intersite exchange term. For a pure XY-type intersite exchange this may be obtained by a Jordan-Wigner transformation. The intuitive argument is that at low temperatures the charge fluctuations in the Kondo lattice model are frozen out and the remaining spin fluctuation spectrum can be simulated by an antiferromagnetic inter-site interaction term of immobile $\tau$ spins coupled by a Kondo interaction to the local noninteracting spins $S$. 

The quantum phase transition between the spin liquid and antiferromagnetic phases
in the one dimensional Kondo necklace model has been studied 
by several methods \cite{jullien,scalettar,moukouri}.
The phase diagram of the genuine Kondo necklace model has been studied by mean field 
approach \cite{Zhang00} in spatial dimension D=1, 2, 3. The one dimensional case is always
in the Kondo singlet phase which is justified by the numerical Monte-Carlo \cite{scalettar} 
and Density Matrix Renormalization Group \cite{moukouri} (DMRG) results. 
While a quantum phase transition between the Kondo singlet phase
and the antiferromagnetic (AF) ordered phase happens by increasing the ratio ($t/J$) of inter-site
to the local exchange interaction for D=2, 3. The effect of both intersite and
intrasite anisotropies on the quantum critical point have also been investigated by the same
approach \cite{langari-thalmeier}. Mean field theory has also been applied to study the
effect of magnetic field on the Kondo necklace model \cite{thalmeier-langari}. However, for D=1 the mean field approach always shows
a non-magnetic Kondo singlet phase even in the presence of anisotropy in the
easy axis term \cite{langari-thalmeier}.

In the present work we want to study the possible quantum phase
transition in the D = 1 Kondo-necklace  model under the
assumption of the anisotropy in the XY-interaction ($\eta$)  between the spin of
itinerant electrons.  However, 
the real space renormalization group (RG) study \cite{saguia}
shows that if the anisotropy in the XY-interaction is greater than some nonzero value
a phase transition from the Kondo singlet to the antiferromagnetic  long 
range order takes place by changing the local exchange coupling ($J/t$). 
It means that for $\eta < \eta_c$ the system is always in Kondo singlet state but for 
 $\eta > \eta_c$ there is a critical exchange coupling ($J_c/t$) which is the border
between the Kondo singlet and AF long range ordered phases. 
However, we will show that the quantum phase 
transition between the Kondo singlet and AF order exists for any nonzero value of $\eta$. 
Our result is based on the spin wave approach 
which is justified in the absence of local exchange for the anisotropic case.
This is in agreement with the general statement which 
tells us about the universality of this transition. Moreover, the Lanczos numerical
computations justify that our statement is correct.

In Sec.\ref{mean-field} we will explain why the mean field approach fails to produce the
correct results in one spatial dimension and easy axis anisotropy. We then consider a 
more general case of anisotropy in the Hamiltonian in Sec.\ref{hamiltonian}.
We then implement a spin wave theory to the Hamiltonian with anisotropy in Sec.\ref{spinwave} and
show that the local exchange will increase the quantum fluctuations which destroy the
AF ordered state. In Sec.\ref{lanczos} we present our numerical Lanczos results which verifies
our proposed phase diagram. Finally, the conclusion and discussion are presented  in
Sec.\ref{discussion}.

\section{Mean field theory}
\label{mean-field}

In this section we shortly review the mean field approach to the Kondo-necklace model.
We will show that this approach fails to represent the antiferromagnetic phase for
Ising like anisotropy.
Let first consider the following Hamiltonian
\begin{eqnarray}
\mathcal{H}_{\rm AKN} = t \sum_{\langle ij \rangle} \left(
\tau_i^x\tau_j^x+\tau_i^y\tau_j^y+\delta \tau_i^z\tau_j^z
\right) + J \sum_i \tau_i \cdot S_i,
\label{akn}
\end{eqnarray}
where $t$ is the strength of exchange coupling between the spin ($\tau$) of itinerant electrons
and $J$ is exchange coupling between the spin of itinerant electrons and localized 
spins ($S$). The anisotropy parameter is represented by $\delta$.
The bond operator representation \cite{Sachdev90} is used to transform the spin operators in terms
of bosonic operators \cite{langari-thalmeier} ($s, t_x, t_y, t_z$). 
We start from the strong coupling limit
where the mean field theory works well and the model has a Kondo singlet ground state.
For the mean field approach \cite{langari-thalmeier}  we assume a singlet condensation 
$\langle s \rangle=\bar{s}$ and a condensation for one triplet component to induce the
antiferromagnetic order $t_{k,z}=\sqrt{N}\bar{t}\delta_{k,Q}+\eta_{k,z}$,
where
$\eta_{k,z}$ is the quantum fluctuation above the triplet condensation
and $Q=\pi$. 
After performing the Bogoliubov
transformation the Hamiltonian in momentum space becomes

\begin{equation}
\mathcal{H}^{mf}_{AKN} =E_{0}^{AF}+\sum_{k} [\sum_{\alpha=x,y}\omega_{\alpha}(k)
{\tilde t}^{\dagger}_{k,\alpha}{\tilde t}_{k,\alpha}+
\omega_{z}(k)
{\tilde \eta}^{\dagger}_{k,z}{\tilde \eta}_{k,z}]
\label{hmf}
\end{equation}
where $E_0^{AF}$ is the ground state energy, $\omega_{\alpha}(k)$ are the excitation 
energies ($\alpha=x, y, z$) and ${\tilde t}^{\dagger}_{k,x(y)}$, ${\tilde \eta}^{\dagger}_{k, z}$ 
(${\tilde t}_{k,x(y)}$, ${\tilde \eta}^{\dagger}_{k, z}$ ) 
are the bosonic creation (annihilation) operators in the diagonal representation.
(More details can be found in Ref.\onlinecite{langari-thalmeier}).
The excitation energies are
\be
\omega_{\alpha}(k)=\sqrt{d_{\alpha}^2(k)-4f_{\alpha}^2(k)}, 
\hspace{5mm}\alpha=x, y, z.
\label{omega}
\ee
where
\bea
 f_x(k)&=& f_y(k)=\frac{t \bar{s}^2}{4} \gamma(k), \nonumber \\
d_x(k)&=&d_y(k)=\mu+\frac{J}{4}+
\frac{t \bar{s}^2}{2} \gamma(k), \nonumber \\
f_z(k)&=&\frac{t \bar{s}^2}{4} \delta \gamma(k), 
\hspace{1mm}d_z(k)=\mu+\frac{J}{4}+
\frac{t \bar{s}^2}{2} \delta \gamma(k),
\label{fd}
\eea
and $\gamma(k)=\cos(k)$.
The ground state energy is
\begin{equation}
E_{0}^{AF}=E_{0}+N\bar{t}^2\left[\mu+\frac{J}{4}-t\bar{s}^2\delta \right],
\end{equation}
where 
\be
E_0=N\Big((\mu-\frac{3J}{4})\bar{s}^2 -\mu\Big) 
+\frac{1}{2} \sum_{k, \alpha=x,y,z}
\Big(\omega_{\alpha}(k)-d_{\alpha}(k)\Big).
\ee
The chemical potential ($\mu$) has been added as a Lagrange multiplier to the
mean field Hamiltonian to 
preserve the dimension of Hilbert space on each bond.
The ground state energy should be minimized with respect to $\bar{s}$, $\bar{t}$
and  $\mu$.  It gives the following equations
\bea
&&\mu=t \delta \bar{s}^2 - \frac{J}{4}, \no \\
&&\bar{t}^2=\frac{5}{4}-\frac{J}{2 t\delta }-
\frac{1}{4N}\sum_{k}\Big(\frac{1}{\sqrt{1+\gamma(k)}}
+\frac{2}{\sqrt{1+\frac{\gamma(k)}{\delta}}}\Big), \no \\
&&\bar{s}^2=\frac{5}{4}+\frac{J}{2 t \delta}
-\frac{1}{4N}\sum_{k}\Big(\sqrt{1+\gamma(k)}
+2\sqrt{1+\frac{\gamma(k)}{\delta}} \Big), \no \\
\label{muts}
\eea
which should be solved self consistently. For $\delta > 1$, we expect to have the long range
antiferromagnetic order in $z$ direction. However, the summation 
$\frac{1}{N}\sum_{k}\frac{1}{\sqrt{1+\gamma(k)}}$ in Eq.(\ref{muts}) diverges which shows that the
mean field approach fails to work correctly in this case.

We have even considered the extreme case of only an Ising term in the interaction
between itinerant spins, namely $t\rightarrow 0$ and $t \delta = constant$.
It is clear that for $J=0$ the ground state is long ranged antiferromagnetic ordered. However,
the above mean field theory fails to show the nonzero AF ordering.
We conclude that the mean field theory is not suitable to show the long ranged
antiferromagnetic phase in one dimension. It should be related to the 
strong quantum fluctuations which have dominant effect in one dimension.
Moreover, the above mean field theory starts from the strong coupling limit ($t=0$) where 
the singlet condensation is supposed to be the case while for the antiferromagnetic
phase we need to consider the weak coupling limit ($J=0$). Having this in mind, we start 
from the weak coupling limit with an AF ordered ground state for the Ising anisotropy case.
The effect of nonzero local exchange ($J\neq0$) increases the quantum fluctuations which
finally destroy the AF long range order. In this respect, we apply the spin wave theory 
to the anisotropic Kondo necklace chain.


\section{The model Hamiltonian}
\label{hamiltonian}
The Hamiltonian defined in Eq.(\ref{akn}) is always in the Kondo singlet phase for
XY anisotropy ($\delta \leq 1$) while it is expected to have a quantum phase transition to
the AF ordered state at a critical exchange coupling ($\frac{J}{t}|_c$) for Ising 
anisotropy ($\delta > 1$). To consider 
a general case we will
investigate the phase diagram of the following Hamiltonian in one dimension
\begin{equation}
\mathcal{H}=t\sum_{i=1}^{N}(\tau_i^z \tau_{i+1}^z +(1-\eta)\tau_i^y \tau_{i+1}^y)+J\sum_{i=1}^{N}(\overrightarrow{\tau_i} \cdot \overrightarrow{S_{i}}),
\label{ham}
\end{equation}
where 
$\eta$ is the anisotropy parameter which defines the deviation from the Ising limit ($\eta=1$). 
The main reason to consider the anisotropy in this form is that  the anisotropy in the itinerant part of the interaction enables us to investigate the effect of symmetry on the results of the Kondo necklace model compared with the Kondo model. 
The present model, the case of $\eta=0$ has U(1) symmetry and for $\eta \neq 0$ it has Z2 symmetry. 
In the case  of full anisotropy ($\eta=1$, Ising limit) 
we expect to have a quantum phase transition  from the antiferromagnetic long range 
order to the Kondo singlet phase. Moreover, in the absence of local exchange coupling ($J=0$)
the universal behavior of the XY term is the same as for all $\eta \neq 0$. It is also another
motivation to see what is the effect of $\eta$ anisotropy on the phase diagram of the model.
Specially, our final results show that there exist an antiferromagnetic phase for
any nonzero value of  $\eta$ when $J$ is smaller than the critical exchange coupling $J_c(\eta)$.
The latter result is in contradiction with the renormalization group result presented
in Ref.\onlinecite{saguia}.

\begin{figure}[ht]
\includegraphics[width=8cm]{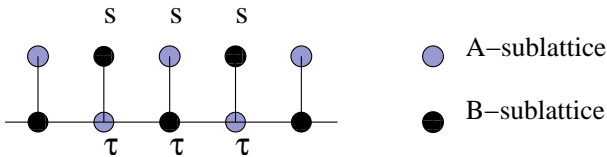}
\caption{\label{fig1} (Color online) One dimensional Kondo-necklace lattice labeled by two sublattices.}
\end{figure}

\section{Spin wave approach} 
\label{spinwave}

The spin wave theory is constructed on a ground state which is maximally polarized.
The low-lying excitations are expected to be created by flipping a spin 
and letting this disturbance to propagate in the crystal which is a spin wave.
These excitations are called magnons which are bosonic quasi particles.

To apply the spin wave approach the one dimensional Kondo-necklace lattice
is labeled by two sublattices A and B (see Fig.\ref{fig1}). The Hamiltonian (\ref{ham})
can be written in the following form where $\sum_{\sigma}$ represents the summation
over the nearest neighbors
\begin{eqnarray}
\mathcal{H}=t\sum_{i=1}^{N/2}\sum_{\sigma}(\tau_i^{zA} \tau_{i+\sigma}^{zB} +(1-\eta)\tau_i^{yA} \tau_{i+\sigma}^{yB}) \nonumber \\
+J\sum_{i=1}^{N/2}(\overrightarrow{\tau_i}^{A} \cdot \overrightarrow{S_{i}}^{B}+
\overrightarrow{\tau_i}^{B} \cdot \overrightarrow{S_{i}}^{A}).
\end{eqnarray}
We then implement the 
Holstein-Primakov transformation \cite{hp} to write the spin operators in 
terms of bosonic operators. Starting from the weak coupling regime ($J=0$) 
the polarized ground state is the Neel state which impose the
following transformations for spins of the itinerant electrons 
\begin{eqnarray}
\label{hpt1}
\tau_i^{zA}=\tau- a_{i}^{\dagger}a_{i}
\hspace{5mm},\hspace{5mm}
\tau_i^{zB}=-\tau+ a_{i}^{\dagger}a_{i}, \\
\tau_{i}^{+A}=(\sqrt{2\tau})(1- \frac{a_i^{\dagger}a_{i}}{2\tau})^{1/2}a_{i},\nonumber \\
\tau_{i}^{+B}=(\sqrt{2\tau})a_{i}^{\dagger}(1- \frac{a_i^{\dagger}a_{i}}{2\tau})^{1/2},
\nonumber \\
\tau_{i}^{-A}=(\sqrt{2\tau})a_{i}^{\dagger}(1- \frac{a_i^{\dagger}a_{i}}{2\tau})^{1/2},\nonumber \\
\tau_{i}^{-B}=(\sqrt{2\tau})(1- \frac{a_i^{\dagger}a_{i}}{2\tau})^{1/2} a_{i}. \nonumber
\end{eqnarray}
In the above equations $\tau$ is the magnitude of spin and 
the Boson creation ($a^{\dagger}$) and annihilation ($a$)
operators obey the following commutation relations,
\begin{equation}
\label{hpt4}
 [a_{i}^{\dagger},a_{j}]=\delta_{i,j}\hspace{2mm},\hspace{2mm}
 [a_{i}^{\dagger},a_{j}^{\dagger}]=0\hspace{2mm},\hspace{2mm}[a_{i},a_{j}]=0.
\end{equation}
A similar transformations is also defined for the localized spin ($S$) by
replacing $\tau \longrightarrow S$ and $a \longrightarrow e$ in Eqs.(\ref{hpt1})
where $e^{\dagger}$ ($e$) represents the creation (annihilation) Boson operator
for the localized spins. Generally, one should define two types of Boson operators
(in each sublattice)
for both itinerant and localized spins. However, due to the translational invariance
symmetry of the model the Boson operators for itinerant spins in each sublattice define
the same operator as defined in Eqs.(\ref{hpt1}). The same story exists also for
the localized ones.

The linear spin wave theory is implemented here where the spin operators are described 
in linear form of the Boson operators and higher order terms have been neglected,
\begin{equation}
\label{lspw}
\tau_{j}^{yA}\simeq\frac{\sqrt{2\tau}}{2i}(a_{j}-a_j^{\dagger})
\hspace{2mm},\hspace{2mm}
\tau_{j}^{yB}\simeq\frac{\sqrt{2\tau}}{2i}(a_j^{\dagger}-a_{j}).
\end{equation}
Similar expressions to Eq.(\ref{lspw}) are applied for the other component 
of spin operators where they are replaced by Boson ones within the 
linear spin wave approximation which finally leads to the following form 
for the whole Hamiltonian 
\begin{eqnarray}
&&\mathcal{H}=-N (t\tau^{2} +J \tau S)+
t\tau\sum_{i=1}^{N/2}\sum_{\sigma}(2 a_i^{\dagger}a_{i}+a_{i+\sigma}^{\dagger}a_{i+\sigma})
\nonumber \\
&-&\frac{t\tau(1-\eta)}{2}\sum_{i=1}^{N/2}\sum_{\sigma}
(a_{i}a_{i+\sigma}^{\dagger}+a_i^{\dagger}a_{i+\sigma}-a_i^{\dagger}a_{i+\sigma}^{\dagger}-a_{i}a_{i+\sigma})
\nonumber \\
&+&2J\sum_{i=1}^{N/2}(S a_i^{\dagger}a_{i}+\tau e_i^{\dagger}e_{i})
+2J\sqrt{\tau S}\sum_{i=1}^{N/2}(a_{i}e_{i}+a^{\dagger}_{i}e^{\dagger}_{i}).
\label{boson-ham}
\end{eqnarray}
To proceed further and diagonalize the Hamiltonian we first transform the 
operators to their Fourier counterparts by the following relations,
\begin{equation}
\label{f1}
a_{j}=\sqrt{\frac{2}{N}}\sum_{k}e^{ik j}c_{k} 
\hspace{5mm},\hspace{5mm}
c_k=\sqrt{\frac{2}{N}}\sum_{j}e^{-ik j} a_j,
\end{equation}
\begin{equation}
\label{f2}
e_{j}=\sqrt{\frac{2}{N}}\sum_{k}e^{ik j}h_{k} 
\hspace{5mm},\hspace{5mm}
h_{k}=\sqrt{\frac{2}{N}}\sum_{j}e^{-ik j}e_{j}.
\end{equation}
In the momentum space representation and for $\tau=s=1/2$ the 
Hamiltonian  (Eq.(\ref{boson-ham})) is  finally written 
in the following form
\begin{eqnarray}
\label{ham-k}
&&\mathcal{H}=-\frac{N}{4}(1+J)
+\frac{(1-\eta)}{4}\sum_{k}\gamma(k)[c_{k}^{\dagger}c_{-k}^{\dagger}+c_{k}c_{-k}] \nonumber \\
&&+\frac{J}{2}\sum_{k}\Big(c_{k}h_{-k}+c_{-k}^{\dagger}h_{k}^{\dagger}+c_{-k}h_{k}\nonumber \\
&&\hspace{35mm}+c_{k}^{\dagger}h_{-k}^{\dagger}+h_{k}^{\dagger}h_{k}+h_{-k}^{\dagger}h_{-k}\Big)
 \nonumber \\
&&+\sum_{k}[(J/2+1)-\frac{(1-\eta)}{4}\gamma(k)][c_{k}^{\dagger}c_{k}+c_{-k}^{\dagger}c_{-k}].
\end{eqnarray}
The Hamiltonian can be diagonalized by the para-unitary transformation \cite{paraunitary}
for the general case of $\eta\neq1$. However, for the case of 
$\eta=1$, one can use the Bogoliubov transformation which is a special case 
of the general para-unitary transformation.


\subsection{$\eta=1$}
 
The Hamiltonian of the general case $\eta\neq1$ can be diagonalized by
the para-unitary transformation \cite{paraunitary} for the bosonic operators.
In this formalism the Hamiltonian is written as a hermitian 2m-square matrix $D$.
The para-unitary transformation ($\Gamma$) can be constructed if and only if
$D$ is positive definite. This procedure is presented in appendix. 

However,
for the special case $\eta=1$  the para-unitary transformation 
is the usual Boguliubov transformation \cite{bogolioubov}.
In this case, the bosonic operators is transformed 
to a new set of bosonic operators ($\alpha, \beta$) by the following relations,
\begin{equation}
\label{bgtransformation}
c_k=u_k \alpha_k + v_k \beta^{\dagger}_{-k} 
\hspace{2mm},\hspace{2mm}
h_k=u_k \beta_k + v_k \alpha^{\dagger}_{-k} 
\end{equation}
where $u_k=\cosh(\theta_k)$ and $v_k=\sinh(\theta_k)$  and $\theta_k$ is the free 
parameter of transformation. The parameter $\theta_k$
should satisfy the following relation to have a diagonal Hamiltonian
\begin{equation}
 \mbox{tanh}(2 \theta_k)=\frac{-J}{J+1},
\end{equation}
which is actually independent of $k$.
The diagonalized Hamiltonian in terms of the new sets of operators is 
\begin{equation}
\mathcal{H}=E_0(J)+\sum_k (\overline{\omega}_{\alpha}(k) \alpha_k^{\dagger} \alpha_k +
\overline{\omega}_{\beta}(k) \beta_k^{\dagger} \beta_k)
\end{equation}
where the ground state energy is 
\begin{equation}
 E_0(J)=N[-\frac{5}{4}(1+J)+\sqrt{1+2J}],
\label{gse-swt}
\end{equation}
and the quasiparticle excitations are
\begin{equation}
 \overline{\omega}_{\alpha}(k)=-1+\sqrt{1+2J}
\hspace{2mm},\hspace{2mm}
 \overline{\omega}_{\beta}(k)=1+\sqrt{1+2J}.
\end{equation}

The sublattice magnetization 
$m_A=\frac{2}{N}\sum_i \langle  \tau-a_{i}^{\dagger}a_{i}\rangle$ defines
the order parameter for the AF long range ordered phase which can be calculated 
in the diagonal bases of the Hamiltonian. For simplicity we put $t=1$ as the scale of energy. 
In the weak coupling limit $J=0$ the Hamiltonian  (Eq.(\ref{ham})) has long range
AF order. The quantum fluctuations which are created by nonzero $J$ decrease the
amount of AF ordering and finally destroy the ordered phase at a critical value, $J_c$,
where the sublattice magnetization becomes zero 
\begin{equation}
\label{avg}
m_A=\frac{1}{2} -\langle a_{i}^{\dagger}a_{i}\rangle=1-\frac{J+1}{2\sqrt{1+2J}}=0
\Longrightarrow J_c\simeq6.46.
\end{equation}
It is the first estimate of the critical value of the exchange coupling.
However, the condition which we 
have stated is the extreme value where the quantum fluctuations suppress the sublattice
magnetization totally. At this extreme condition the accuracy of the linear spin
wave results is not justified, since the amplitude of the quantum fluctuations 
is not small. However, this approach is valid for small amplitude of fluctuations.
Moreover, it shows that the amplitude of fluctuations become large by increasing the
value of exchange coupling ($J$). Thus, the qualitative picture of the spin wave
approximation is correct. 

\begin{figure}[ht]
\includegraphics[width=8cm]{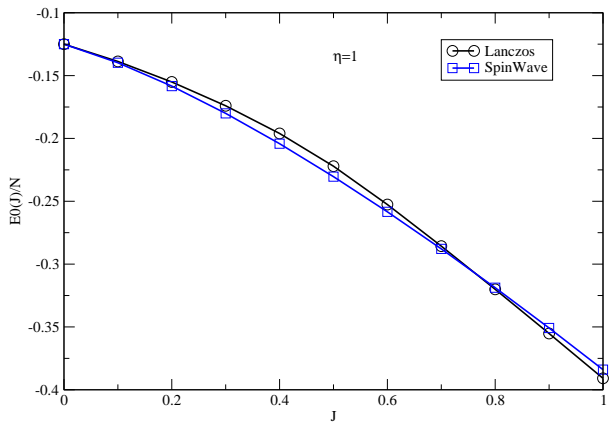}
\caption{\label{fig2} (Color online)
Ground state energy per site ($E_0(J)/N$) versus $J$ for $\eta=1$ 
obtained by numerical Lanczos method (extrapolated to the $N\rightarrow \infty$ with 
4 digits accuracy)
and linear spin wave theory which show very good agreement.
}
\end{figure}

To justify that the spin wave theory gives correct results, we have plotted the 
ground state energy per site versus the exchange coupling ($J$) in Fig.\ref{fig2}.
We have compared the results obtained in Eq.(\ref{gse-swt}) with the numerical Lanczos
ones which will be discussed in the next section. It shows very good agreement of the
spin wave with Lanczos results which justifies that the spin wave ground state presented
here can be a good candidate to represent the many body ground state in this case.

\subsection{$\eta\neq1$}
We have implemented the para-unitary transformation (see appendix)
to diagonalize the Hamiltonian (Eq.(\ref{ham-k})). The diagonalized Hamiltonian
in terms of new sets of bosonic operators ($\phi, \psi$) is 
\begin{equation}
\label{ham-k-d}
\mathcal{H}=E_0(\eta, J)+\sum_k [\Omega_{\alpha}(k) \phi_k^{\dagger} \phi_k +
\Omega_{\beta}(k) \psi_k^{\dagger} \psi_k].
\end{equation}
where $E_0(\eta, J)=(1/N)\sum_k e_0(\eta, J; k)$ is the ground state energy 
and  $e_0(\eta, J; k)$ has the following expression
\bea
e_0(\eta, J; k)=\frac{1}{2\sqrt{2}}
\big( \sqrt{\mathcal{P}-\mathcal{Q}}+\sqrt{\mathcal{P}+\mathcal{Q}}\big), \no \\
\mathcal{P}=4(1+J)-(2+J)\gamma(k) \;\;,\;\; \mathcal{Q}=\sqrt{\mathcal{S}+\mathcal{T}^2},\no \\
\mathcal{S}=(2+J)\gamma(k) \eta -8 J^2 (2+\gamma(k)(\eta-1)), \no \\
\mathcal{T}=4+J(4+\gamma(k)(\eta-1))+2\gamma(k)(\eta-1).
\eea

The frequencies $\Omega_{\alpha}(k)$ and $\Omega_{\beta}(k)$ are 
\bea
\Omega_{\alpha}(k)=\frac{1}{\sqrt{2}}\sqrt{4(1+J)-(2+J)\gamma(k)-\mathcal{A}} \no \\
\Omega_{\beta}(k)=\frac{1}{\sqrt{2}}\sqrt{4(1+J)-(2+J)\gamma(k)+\mathcal{A}},
\eea
where
\begin{eqnarray}
\mathcal{A}=\sqrt{\mathcal{S}+\mathcal{T}^2}.
\end{eqnarray}
A remark is in order here, the results in this subsection will not reproduce the
results of $\eta=1$ case simply by putting $\eta$ equal to one. The reason is related to
a divergent factor of $(1-\eta)^{-1}$ which appears in the intermediate part of calculations.
Such terms will not appear if we put $\eta=1$ from the beginning and the calculations will
become much simpler as presented in the last subsection.
In the diagonal bases of the Hamiltonian we can calculate the sublattice magnetization
as defined in Eq.(\ref{hpt1}). As we have discussed in the previous subsection we can 
determine the critical border between the AF order and the Kondo singlet phase by looking for
the position where the sublattice magnetization becomes zero.

Our results which are summerized in Table.\ref{t1}. show that the onset of nonzero
XY-anisotropy ($\eta\neq0$) develops the long range AF order for small local
exchange coupling ($J<J_c$). This is also justified by the numerical Lanczos results where
the static structure factor at momentum $\pi$ diverges for 
any nonzero $\eta$ and $J<J_c$. However, the numerical values for $J_c$ which are obtained
by the spin wave approach is far from the numerical Lanczos ones. This will be discussed in 
the Sec.\ref{discussion}.


\section{Numerical Lanczos Results }
\label{lanczos}

The numerical Lanczos method has been applied to the one dimensional anisotropic Kondo-necklace
model defined in Eq.(\ref{ham}). The anisotropic Hamiltonian does not commute with the
total $z$-component spin. Thus, one should consider the full Hilbert space for doing numerics.
In this respect, we have performed the numerical computations with $N=12, 16, 20, 24$ spins.
The computations have been done for different values of the exchange coupling ($J$) and
the anisotropy parameter ($\eta$) to trace the quantum phase transition between the Kondo singlet
phase and the antiferromagnetic one. The Kondo singlet ground state is not ordered while the
antiferromagnetic phase has long range order. The long range order imposes that the structure
factor diverges in the thermodynamic limit ($N\rightarrow\infty$) for the antiferromagnetic
wave vector $k=\pi$. The structure factor of the $z$-component spin at the wave vector
$k$ is defined by
\begin{equation}
G(k)=\sum_{r=0}^{\frac{N}{2}-1} \la \tau_0^z \tau_{r}^z \ra e^{i k r}.
\label{stf}
\end{equation}

\begin{figure}[ht]
\includegraphics[width=8cm]{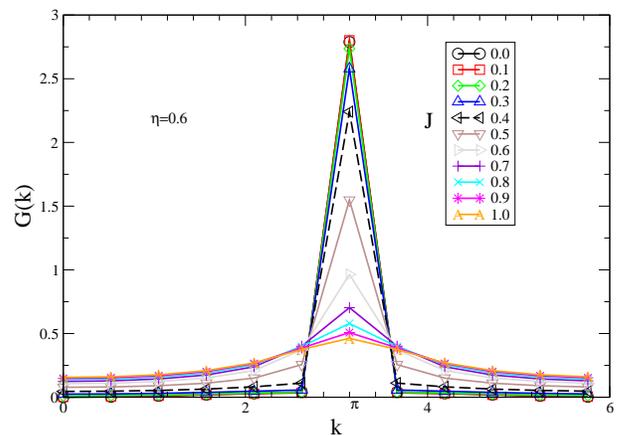}
\caption{\label{fig3} (Color online)
The $z$-component structure factor versus $k$ for $\eta=0.6$, $N=24$ and
different $J=0.0, 0.1, 0.2, \dots, 1.0$.
}
\end{figure}
We have plotted the structure factor ($G(k)$) versus the wave vector ($k$) for
$\eta=0.6$, $N=24$ and different exchange couplings ($J=0.0, 0.1, 0.2, \dots, 1.0$)
in Fig.\ref{fig3}. The structure factor gets a strong peak at $k=\pi$ for some values
of $J$. This peak shows the presence of antiferromagnetic order. We will show in the
next plots how the height of this peak evolve as the size of system changes.

\begin{figure}[ht]
\includegraphics[width=8cm]{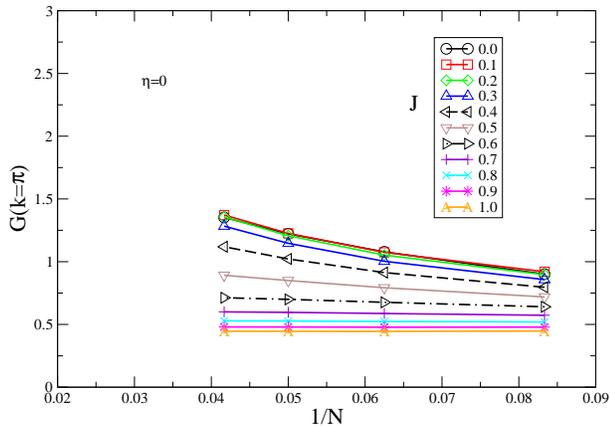}
\caption{\label{fig4} (Color online)
The $z$-component structure factor for $k=\pi$ versus the inverse of 
of system size ($1/N$) for $\eta=0.0$ and
different $J=0.0, 0.1, 0.2, \dots, 1.0$. The structure factor diverges for $J=0$
like $\sqrt{N}$ which represent the massless behavior of XX spin 1/2 chain.
The other sets of data show no divergence or weaker than $\sqrt{N}$ which justifies
no long range antiferromagnetic order.
}
\end{figure}

\begin{figure}[ht]
\includegraphics[width=8cm]{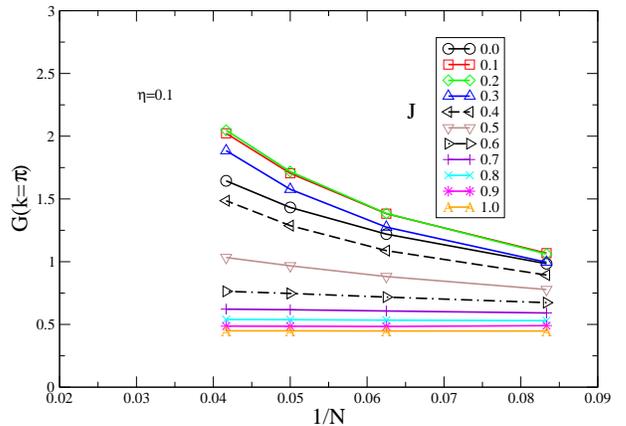}
\caption{\label{fig5} (Color online)
The $z$-component structure factor versus $1/N$ (inverse of size of system) 
at $k=\pi$ for $\eta=0.1$ and
different $J=0.0, 0.1, 0.2, \dots, 1.0$. The divergence of $G(k=\pi)$ 
as $\frac{1}{N}\rightarrow 0$ for $J\leq 0.3$ is faster than $\sqrt{N}$. 
It shows the antiferromagnetic ordered phase
for $J\leq 0.3$.
}
\end{figure}

\begin{figure}[ht]
\includegraphics[width=8cm]{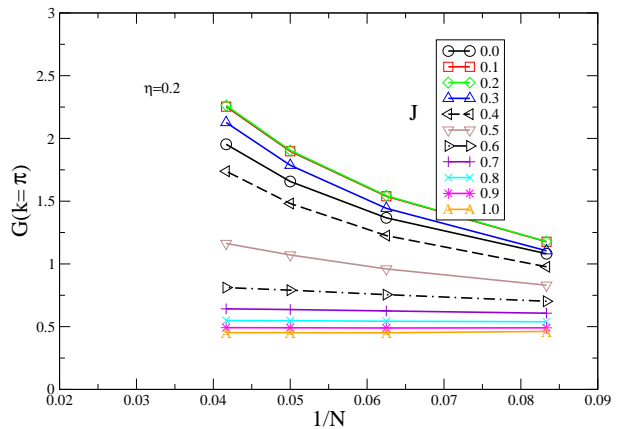}
\caption{\label{fig6} (Color online)
$G(k=\pi)$ (the $z$-component structure factor) versus $1/N$ for $\eta=0.2$ and
different $J=0.0, 0.1, 0.2, \dots, 1.0$. It is clear that $G(k=\pi)$ diverges
for $J<0.4$ as $N$ goes to infinity.
}
\end{figure}

\begin{figure}[ht]
\includegraphics[width=8cm]{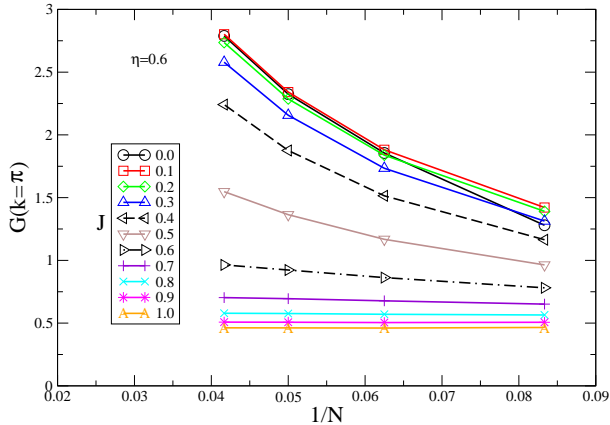}
\caption{\label{fig7} (Color online)
The $z$-component structure factor versus $1/N$ for $k=\pi$ and $\eta=0.6$ and
different $J=0.0, 0.1, 0.2, \dots, 1.0$. The structure factor diverges
for $J\leq0.4$ and $\frac{1}{N}\rightarrow 0$.
}
\end{figure}

\begin{figure}[ht]
\includegraphics[width=8cm]{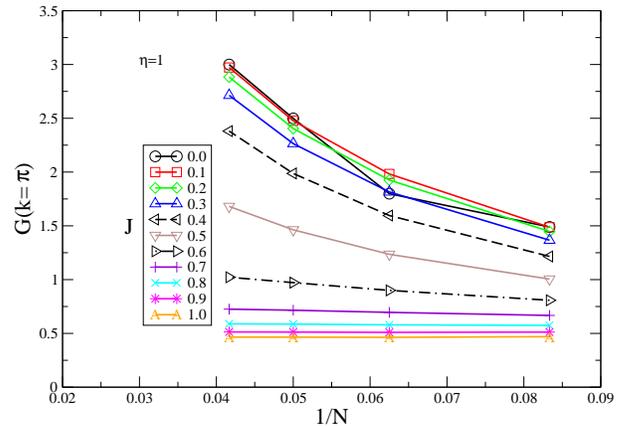}
\caption{\label{fig8} (Color online)
The $z$-component structure factor versus $1/N$ at $k=\pi$ for $\eta=1.0$ and
different $J=0.0, 0.1, 0.2, \dots, 1.0$. The divergence of $G(k=\pi)$ for 
$N\rightarrow\infty$ is clear for $J<0.5$.
}
\end{figure}

The $z$-component spin of structure factor at the antiferromagnetic 
wave vector, $k=\pi$, has been plotted in Figs.\ref{fig4}-\ref{fig8}
for different anisotropy ($\eta$) versus the inverse of system size ($N$).
The system sizes are $N=12, 16, 20, 24$.
Each figure contains different plots for $J=0.0, 0.1, 0.2, \dots, 1.0$.

We have plotted the case of genuine  Kondo-necklace model, $\eta=0$ in Fig.\ref{fig4}.
For $J=0$, the properties of the XX spin 1/2 chain is reproduced which
shows that the structure factor diverges \cite{karbach} proportional to $\sqrt{N}$.
However, it does not show the antiferromagnetic order of this model. It only represent
the critical properties of the XX spin 1/2 chain which has algebraic decay of correlation
functions. Turning on the local exchange to nonzero values $J\neq0$ we observe weaker growth
of the structure factor as the system size in increased. It is concluded that the 
one dimensional genuine Kondo-necklace model does not have an antiferromagnetic phase and
is always in the Kondo singlet state. Moreover, we expect 
the structure factor diverges stronger than $\sqrt{N}$ for 
the antiferromagnetic ordered phase.

The presence of anisotropy ($\eta\neq0$) breaks the U(1) symmetry 
and reduces it to Z2 symmetry. This is the onset of antiferromagnetic
ordering formation which is the result of symmetry breaking. The structure 
factor at $k=\pi$ versus the inverse of length ($1/N$) for $\eta=0.1$ has 
been plotted in Fig.\ref{fig5}. It is clear that $G(\pi)$ diverges 
faster than $\sqrt{N}$ for $J=0$ which verifies the formation of 
antiferromagnetic ordering in the ground state. A similar situation 
exists also for $J=0.1, 0.2, 0.3, 0.4$. However, for $J\ge 0.5$ the divergence
of $G(\pi)$ is not apparent as $N\rightarrow \infty$. We then conclude that
for $J\le J_c(\eta=0.1)\simeq 0.4$ the model has antiferromagnetic ordering while
for larger values of $J$ it is in the Kondo singlet phase. Thus, a quantum phase
transition occurs at $J_c$ from the antiferromagnetic order to the disordered Kondo singlet
state.

We have also plotted the z-component structure factor at antiferromagnetic
wave length ($G(k=\pi)$) versus the inverse of system size ($1/N$)
for $\eta=0.2$ in Fig.\ref{fig6}, $\eta=0.6$ in Fig.\ref{fig7} and
$\eta=1.0$ in Fig.\ref{fig8}. We have observed that  $G(k=\pi)$
diverges in the thermodynamic limit ($N\rightarrow \infty$) for 
$J\leq J_c(\eta)$ which verifies the existence of antiferromagnetic ordering.
Our results for the critical exchange have been summerized in table.\ref{t1}.
The accuracy of our results for the critical point ($J_c$) is about $\pm 0.05$.
Because, the size of system is limited to $N=24$. To distinguish between
the ordered and disordered phase more accurately it is necessary to consider larger 
system sizes. 

\begin{table}
\caption{\label{t1} The critical exchange value ($J_c$) 
for different values of anisotropy ($\eta$). The Lanczos results
are read from the data of $G(\pi)$ versus $1/N$ which has an 
accuracy about $\pm 0.05$. }
\begin{ruledtabular}
\begin{tabular}{ccccc}
$\eta$&0.1&0.2&0.6&1.0\\
\hline
$J_c$ (Lanczos)&0.3&0.35&0.4&0.45\\
\hline
$J_c$ (Spin Wave)&6.13&6.21&6.39&6.46\\
\end{tabular}
\end{ruledtabular}
\end{table}

\section{Conclusion and Discussion}
\label{discussion}

We have studied the phase diagram of the one dimensional anisotropic
Kondo necklace model. We have focused our attention on the case of
Ising anisotropy (easy axis of the spin of itinerant electrons, $\delta >1$). Let us suppose that
the interaction between the spin of itinerant electrons is only 
composed of $\tau^z_i \tau^z_j$ type. It is clear that in the absence of 
local exchange interaction ($J=0$) the model has a classical antiferromagnetic
ordered ground state with nonzero staggered magnetization. It is expected that a
nonzero local exchange ($J\neq0$) adds quantum fluctuations which destroy the
antiferromagnetic order at a critical exchange ($J_c$). However, the mean field
approach \cite{langari-thalmeier} fails to produce this picture correctly. In the
mean field approach which is based on the strong coupling limit ($J/t \rightarrow \infty$) 
we have assumed a nonzero condensation for local singlet formation
plus a nonzero triplet occupation ($\bar{t}$) which induces the antiferromagnetic order. 
The solution of the self consistent mean field equations always give no magnetic
order, $\bar{t}=0$. A divergent integral appears if we assume $\bar{t}\neq0$.
However, the mean filed approach works fairly well in two and three dimensional models
in addition to the case of XY anisotropy ($0<\delta <1$) of the 
one dimensional system \cite{langari-thalmeier}.

\begin{figure}[ht]
\includegraphics[width=8cm]{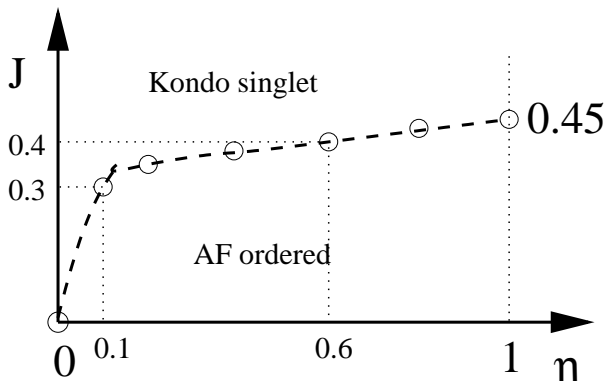}
\caption{\label{fig9} 
Phase diagram of the one dimensional anisotropic Kondo necklace model where a
quantum phase transition (dashed line) separates the ordered antiferromagnetic
phase from the Kondo singlet one.}
\end{figure}

To overcome the problem which appears in the mean field approach for Ising
anisotropy of the one dimensional model we decided to start from the weak coupling
limit ($J\rightarrow 0$). The linear spin wave theory has been implemented which
is based on a classical antiferromagnetic ground state. To study the phase diagram 
we have considered a general case of Hamiltonian which has been given in Eq.(\ref{ham}).
Our results show that the quantum fluctuations grow with increasing $J$ such that
the sublattice magnetization becomes zero at a critical exchange coupling ($J_c$).
Moreover, the antiferromagnetic phase exists for any nonzero value of anisotropy 
($\eta\neq0$) and $J<J_c$. It is in contradiction with the results presented in 
Ref.\onlinecite{saguia} which gives the antiferromagnetic phase only for $\eta>\eta_c\simeq0.58$.
In other words our results state that $\eta_c=0$. We have plotted the phase diagram 
of the one dimensional anisotropic Kondo necklace model in the $J-\eta$ plane in Fig.\ref{fig9}.
It is obvious that for any nonzero $\eta$ there exists an antiferromagnetic phase. Moreover,
the critical exchange coupling ($J_c$) depends strongly on $\eta$ for small anisotropy,
$0<\eta<0.1$ while its dependence is weak for $\eta>0.1$.

Our result is in agreement with the symmetry arguments which is based on a qualitative
picture. Let us first suppose that $J=0$. The remaining Hamiltonian is an anisotropic XY 
spin 1/2 chain. The quantum renormalization group \cite{langari} verifies that the universality 
class of the nonzero anisotropy ($\eta\neq0$) is the same as Ising case ($\eta=1$). Thus, 
if the antiferromagnetic phase exists for the Ising case it should also appear for any
nonzero value of $0<\eta<1$. 
Moreover, the Hamiltonian
has U(1) symmetry at $\eta=0$ while the symmetry is broken to the lower Z2 for $\eta\neq0$.
It is thus expected that the quantum phase transition happens at the symmetry breaking point. 
We expect that the addition of local exchange ($J\neq0$) becomes irrelevant 
at the fixed point $J=0$, which is in agreement with our numerical results.
However, this needs to be proved by more sophisticated methods like
weak-coupling renormalization group which is out of the scope of this article.

To verify that our result gives the correct phase diagram we have implemented
the numerical Lanczos method. Our results show that the z-component structure factor
diverges at the antiferromagnetic wave vector for any nonzero $\eta$ and $J<J_c$.
However, the value of the quantum critical point ($J_c$) can not be determined with
very high accuracy because of the finite size effect. In this study we have been limited
to $N=24$ because the full Hilbert space should be considered in computation and the 
Lanczos is basically limited to such values. However, it is proposed that a density matrix 
renormalization group can get more accurate values. Moreover, we have understood that the
critical exchange ($J_c$) which is given by spin wave theory is far from the real one which
is given by numerical Lanczos method. The reason is related to the strong quantum fluctuations
which exist close to critical point and ruin the accuracy of linear spin wave results.

\section{Acknowledgments}
This work was supported in part by the center of excellence in Complex
Systems and Condensed Matter (www.cscm.ir). The authors acknowledge the support 
of Sharif University of Technology.

\appendix

\section{The para-unitary transformation}
This section is based on the approach proposed in Ref.\onlinecite{paraunitary}
to diagonalize a bosonic Hamiltonian.
The Hamiltonian in Eq.(\ref{boson-ham}) can be written in the following matrix
form 
\be
\mathcal{H}=\sum_k \Gamma^{\dagger} D \Gamma
\label{a1}
\ee
where $\Gamma^{\dagger}$ is a row vector 
\be
\Gamma^{\dagger}=(c^{\dagger}_k \hspace{2mm} h^{\dagger}_k \hspace{2mm} c_{-k} \hspace{2mm} h_{-k}),
\label{a2}
\ee
and $D$ is a square matrix
\bea
 D=\left(\begin{array}{cccc} \frac{2J+4-(1-\eta)\gamma(k)}{4}
&0&\frac{(1-\eta)\gamma(k)}{4}&\frac{J}{2}\\
0&\frac{J}{2}&\frac{J}{2}&0\\
\frac{(1-\eta)\gamma(k)}{4}&\frac{J}{2}&\frac{2J+4-(1-\eta)\gamma(k)}{4}&0\\
\frac{J}{2}&0&0&\frac{J}{2}
\end{array}\right).
\label{a3}
\eea

The Hamiltonian will be diagonalized by the following transformation 
($\mathcal{J}$) to 
the new set of boson operators,
\be
\Lambda=\mathcal{J}\Gamma,
\label{a4}
\ee
where 
\be
\Lambda^{\dagger}=(\phi^{\dagger}_k \hspace{2mm} \psi^{\dagger}_k \hspace{2mm} \phi_{-k} \hspace{2mm}
\psi_{-k}),
\label{a5}
\ee
and the diagonal representation is given in Eq.(\ref{ham-k-d}).
The new set of operators should satisfy the bosonic commutation relation, namely
\be
[\Lambda_i, \Lambda^{\dagger}_{j}]=\hat{\delta}_{i, j},
\label{a6}
\ee
where $\hat{\delta}$ is the para Kronecker symbol defined by
\bea
\hat{\delta}_{i,i}=1 \hspace{3mm}&,&\hspace{3mm} 1\leq i \leq 2, \no \\
\hat{\delta}_{i,i}=-1 \hspace{3mm}&,&\hspace{3mm} 3\leq i \leq 4, \no \\
\hat{\delta}_{i,j}=0 \hspace{3mm}&,&\hspace{3mm} i\neq j.
\label{a7}
\eea
In other words the transformation matrix ($\mathcal{J}$) should obey the following 
equations,
\be
\mathcal{J} \hat{I} \mathcal{J}^{\dagger}= \hat{I}
\hspace{3mm} \hbox{or} \hspace{3mm}
\mathcal{J}^{\dagger} \hat{I} \mathcal{J}= \hat{I},
\label{a8}
\ee
where $\hat{I}$ is the para unit matrix, 
$\hat{I} \equiv diag (1, 1, -1, -1)$.
A theorem which has been proved in Ref.\onlinecite{paraunitary} states that
the matrix $D$ can be para unitary diagonalized  into a matrix with
all diagonal elements positive if and only if $D$ is positive definite. 
According to Eq.(\ref{a3}), $D$ is positive definite except at the point ($\eta=0$,
$J=0$). To find the para unitary transformation which satisfies Eq.(\ref{a8}) the
Hamiltonian and the transformation will be written in the following matrix 
form
\bea
 D=\left(\begin{array}{cc} A
&B\\
B&A
\end{array}\right).
\label{a9}
\eea
and 
\bea
 \mathcal{J}=\left(\begin{array}{cc} U^{\dagger}
&-V^{\dagger}\\
-V^{\dagger}&U^{\dagger}
\end{array}\right).
\label{a10}
\eea
where $A, B, U$ and $V$ are $2\times2$ matrices.

The $2\times2$ matrix $K$ is defined by $K^{\dagger} K=A-B$. 
The eigenvalues of the Hermitian matrix $K(A+B)K^{\dagger}$ are
identified by $\mbox{det}[K(A+B)K^{\dagger} - \lambda^2_i I]=0$, 
$i=1,2$  where $I$ is the  $2\times2$ unit matrix and the corresponding eigenvectors ($\chi_i$)
are normalized such that $\chi_i^{\dagger} \chi_i=\frac{1}{\lambda_i}$.
We define $f_i=K^{\dagger}\chi_i$ and obtain $p_i$ such that
\be 
(A+B) f_i= \lambda_i p_i.
\label{a11}
\ee
The transformation matrices $U$ and $V$ are given by their column vectors with 
the following relations,
\bea
u_i=\frac{1}{2}(f_i+p_i), \no \\
v_i=\frac{1}{2}(f_i-p_i).
\label{a12}
\eea


\end{document}